\documentclass[fleqn,10pt]{wlscirep}
\usepackage{amsmath}
\usepackage{algorithm}
\usepackage[noend]{algpseudocode}
\usepackage[colorinlistoftodos]{todonotes}
\usepackage{graphicx,graphics,subfigure,multicol} 
\usepackage{hyperref}

\title{Understanding the Spatial and Temporal Activity Patterns of Subway Mobility Flows}

\author[1]{Zhanwei Du}
\author[2]{Bo Yang}
\author[1*]{Jiming Liu}
\affil[1]{Department of Computer Science, Hong Kong Baptist University, Kowloon Tong KLN, Hong Kong}
\affil[2]{School of Computer Science and Technology, Jilin University, Changchun, China}

\affil[*]{jiming@comp.hkbu.edu.hk}

\begin{abstract}
In urban transportation systems, mobility flows in the subway system reflect the spatial and temporal dynamics of working days. To investigate the variability of mobility flows, we analyse the spatial community through a series of snapshots of subway stations over sequential periods. Using Shanghai as a case study, we find that the spatial community snapshots reveal dynamic passenger activities. Adopting a dual-perspective, we apply spatial and temporal models separately to explore where and when individuals travel for entertainment. In the two models, microblog topics and spatial facilities such as food venues and entertainment businesses are used to characterise the spatial popularity of each station and people's travelling perceptions. In the studied case, the city centre is characterised by greater social influence, and it is better described by the spatial model. In the temporal model, shorter travel distances motivate individuals to start their trips earlier. Interestingly, as the number of food-related facilities near the starting station increases, until it exceeds 1563, the speed of people's journeys slows down. 
This study provides a method for modelling the effects of social features on mobility flows and for predicting the spatial-temporal mobility flows of newly built subway stations.

\end{abstract}
\begin{document}

\flushbottom
\maketitle
\thispagestyle{empty}

\section*{Introduction}
Urban transport plays an important role in shaping and reflecting the evolution of cities \cite{batty2013theory,batty2008size}. To create the desired social-economic outputs, urban transport planners use human mobility flows to understand the spatial-temporal interactions of people \cite{batty2013theory} on transportation systems. 
An urban public transport network (UTN) consists of the mobility flows of an urban transportation system, where the nodes represent the public transit stand locations and the directional edges denote the mobility flows from one node to another \cite{rodrigue2009geography}. For example, in a subway network, each node denotes a subway station and each edge denotes the mobility flow between two nodes, weighted by the volume. 

The mobility flows {dynamically} change over time as people's reasons for travelling change {activities} (e.g. work or entertainment) \cite{zhong2015measuring}. 
To measure the {variability of spatial-temporal mobility flows}, {communities} in a UTN (a community is a set of densely interconnected nodes that have few connections to outside nodes \cite{newman2003structure}) are used to show the dynamic changes in UTN over time.
Furthermore, a {community snapshot} is a snapshot of the communities in a UTN at a single point in time (e.g. a period, day or year). Community snapshots have many applications in transport planning, e.g. in urban development analysis, experts use them to quantify the influence of urban development on transportation networks \cite{sun2015quantifying,zhong2014detecting}; in urban dynamics analysis, experts use them to measure the variability of human mobility patterns \cite{zhong2015measuring}; and in urban area analysis, experts use them to identify functional zones \cite{kim2014analysis}. 
The community method is not the only means of observing dynamic mobility flows; {driven spatial-temporal models with high-order structures of activity patterns} are also valuable in many scenarios. For example, in public transit scheduling and pricing, ticket prices are optimised to ease traffic congestion by influencing passengers' driven models of temporal scheduling \cite{de2015discomfort}; in urban planning, cities can be planned according to valuable factors, i.e. factors that attract individuals' decisions to live in a or visit a particular place \cite{paldino2015urban}. 
To further understand the dynamic decision-making processes that shape individuals' spatial-temporal movements, we study dynamic mobility flows by measuring the variability in a heterogeneous sample of community snapshots in a UTN and explore the models that drive these patterns.

 We observe the dynamic mobility flows using community snapshots of different spatial stations over time. These flows exist in many public transit systems (e.g. shared bicycle systems \cite{austwick2013structure,borgnat2013dynamical}, public bus systems \cite{chatterjee2015studies,zhang2015evaluation} and taxi systems \cite{liu2015revealing,kang2013exploring}).
Here, we use a subway system, an important part of a public transport network \cite{rodrigue2009geography}, to demonstrate the use of community snapshots. 
As current static or aggregated mobility measures are robust to perturbations and cannot clearly reveal the variability of temporal community structures \cite{zhong2015measuring}, communities patterns are commonly studied retrospectively to get a deeper understanding of how people's activities change over time. Here, we study changes in the use of a subway system by detecting and comparing community snapshots from different time periods \cite{zhong2015measuring,sun2015quantifying}.
Retrospective studies of community snapshots improve our ability to measure the mobility flows from the perspective of the network. However, they do not provide insight into the {spatial-temporal} models that drive the mobility dynamics. For the current day, we cannot determine where individuals would like to go, let alone when they would like to begin any of their activities. 

To address this issue, we also study the patterns of passengers' movements between places \cite{diao2015inferring,gong2015inferring,alexander2015origin,jiang2012discovering} as a high-order spatial-temporal structure.
The examination of the activity patterns should answer two questions: { 
 (1) where do individuals travel to for their activities; and
 (2) when do people start their activities? }
For example, in the evening, can our models {infer} where and when a person is likely to go for entertainment after work? 
These studies will raise planners' awareness of human movements between stations over different time periods.
Specifically, the likelihood of an individual participating in an activity in a particular spatial-temporal space is associated with both the station's spatial characteristics (such as population density and number of retail venues) and temporal characteristics (such as the day of the week and time of day) \cite{diao2015inferring}.


Previous studies of the {where (spatial) }dimension of intra-urban spatial mobility in public transit have typically focused on regional populations (such as Beijing \cite{liang2013unraveling,yan2014universal}, Shenzhen \cite{yan2014universal}, London \cite{liang2013unraveling}, Chicago \cite{liang2013unraveling,yan2014universal}, Los Angeles \cite{liang2013unraveling} and Abidjan \cite{yan2014universal}) and travelling distance (e.g. in Seoul \cite{goh2012modification}). 
However, the correlation between public transit and its surrounding economic and social environment is obvious (such as in Biscay \cite{mendiola2014link}), especially the subway system (e.g. in Sao Paulo\cite{haddad2015underground}).
Most of the existing studies focus on universal laws of human intra-urban mobility, but neglect the {influence of a heterogeneous spatial environment} on activities over periods of time, and thus ignore the dynamics of universal mobility laws \cite{goh2012modification,perkins2014theory}. 
Thus, 
the {where (spatial) }dimension can be studied by analysing  individuals' spatial movements as revealed by measurements of stations' {popularity in a heterogeneous spatial environment}. 
Static information (such as population) cannot reflect dynamic spatial popularity over periods of time. Some studies have used individual digital traces to detect the urban magnetism of different places (e.g. in New York City \cite{paldino2015urban}). For example, location-based {microblogs} such as Twitter \cite{preoctiuc2015analysis,preoctiuc2015studying} have been found to correlate with individuals' profiles, spatial-temporal behaviour and preferences\cite{yuan2013and}. Thus, in this study, each station's spatial popularity is measured by the volume of spatial microblogs associated with it. 
The study uses a location-based mobility model based on the spatial microblogs of neighbouring stations to describe the {heterogeneous spatial popularity} of each station, i.e. its attractiveness to individuals.

Existing studies of the {when (temporal) }dimension of passengers' activities focus on factors in the traffic system such as trip fares, delay cost and travel distance \cite{mohring1972optimization}, and on travel discomfort or congestion \cite{kraus1991discomfort,de2015analyzing,de2015discomfort}). They use a variety of methods such as the equilibrium equation \cite{kraus1991discomfort,de2015analyzing,de2015discomfort} to measure the effect of these factors.
However, it is not enough to explain the dynamic {uncertainty} in individuals' scheduling decisions, especially under the influence of a particular {social and economic environment}. 
A passenger's travelling objective is to find an equilibrium between travelling comfort and schedule delay cost \cite{de2015discomfort}. Thus, differential equations can be used to model individuals' decision-making processes regarding temporal activities that balance travelling comfort and delay cost. 
More specifically, a station's perceived travelling comfort can be measured by the number of {business buildings} surrounding the station (for Seoul see \cite{bae2003impact} and for Shanghai see \cite{jiwei2006railway}). 
In addition, the perception of delay cost correlates with the distance between two places \cite{de2015discomfort}. Thus distance can be used to infer individuals' perception of delay cost.
\textcolor{black}{ 
In this way, we study the {when (temporal)} dimension of flows by considering the balance between travelling discomfort and delay cost. To characterise individuals' perceptions, features such as spatial facilities are correlated with perceptions through a generalised additive model\cite{hastie1990generalized,wood2006generalized}. 
}

In this study, we take the subway system of Shanghai as a case study. This dataset provides the detailed trace information of 11 million individuals over a one-month period, including check-in and check-out times for each subway trip. The aggregated data on the subway system are collected by the Shanghai Public Transportation Card Co. Ltd, and released by the organising committee of the Shanghai Open Data Apps \cite{coltd2015}. 
To examine the stations' environments, we collect microblog data and information about the spatial facilities around each station from Baidu APIs \cite{baidu_Shanghai} and Weibo APIs\cite{Weibo_Shanghai} for the studied month.

\textcolor{black}{ 
In our case study, we {first} measure the variability of mobility flows by investigating the dynamic spatial-temporal community snapshots of the mobility flows. 
The community snapshots taken at different periods (morning, morning/afternoon and evening) do not agree with each other; the evening snapshots are particularly distinct. 
We further investigate the high-order structure of the evening activity patterns. 
The findings show that most individuals return home after work. Activity patterns with more edges are less common.
{\textit{In addition}}, we use spatial and temporal models to examine the effects of social factors on activity patterns. Specifically, in our examination of {the where dimension}, we find that the city centre has a higher social influence. This influence is better described by the spatial model, which illustrates heterogeneous spatial popularity.
{\textit{Finally}}, in the exploration of {the when dimension}, we find that the individuals tend to start their trips earlier when they travel shorter distances. Interestingly, if there are more food-related facilities (but no more than 1563) near the starting station, people are more likely to slow down their trip to avoid travelling discomfort.
}
Our results deepen the understanding of {spatial-temporal} mobility flows in urban public transport networks by helping to model and estimate the spatial-temporal mobility flows. Specifically, we highlight the effects of social influences as measured by microblogs and spatial facilities. 

\begin{figure}[th]
\centering \includegraphics[scale=1.60]{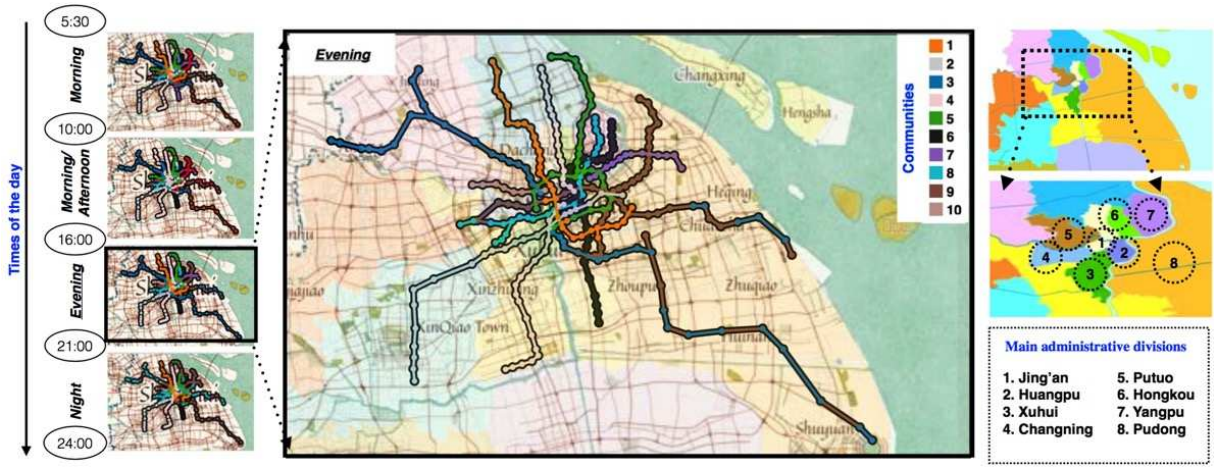}
\caption{{Snapshots of communities at different times, embedded in geographical regions.} 
\textcolor{black}{Taking one working day as an example, these snapshots of communities at four different times of day \textit{{ (morning, morning/afternoon, evening and night)} show the structure of the subway system.} 
The various communities are indicated by the colour of the nodes in the subway system. 
A comparison of  any two adjacent periods shows that the communities move, as shown by the shifting nodes, especially in the evening and at night. Due to the low passenger volume at night, we focus on the evening period. An extended evening snapshot, shown in the middle of the diagram, combines the eight administrative divisions of Shanghai \cite{wiki_Shanghai}. 
The dynamic community snapshots reveal passengers' various decisions about where and when they travel for different activities over a one-day period. These data are useful for exploring the where and when dimensions of mobility flows. 
The spatial map was created using OpenStreetMap online platform (\href{http://http://www.openstreetmap.org/}{http://http://www.openstreetmap.org/}) (© OpenStreetMap contributors) under the license of CC BY-SA (\href{http://www.openstreetmap.org/copyright}{http://www.openstreetmap.org/copyright}). More details of the licence can be found in  \href{http://creativecommons.org/licenses/by-sa/2.0/}{http://creativecommons.org/licenses/by-sa/2.0/}.
Line graphs were drawn using Tableau Software for Desktop version 9.2.15 (\href{https://www.tableau.com/zh-cn/support/releases/9.2.15}{https://www.tableau.com/zh-cn/support/releases/9.2.15}).
The layouts were modified with Keynote version 6.6.2 (\href{http://www.apple.com/keynote/}{http://www.apple.com/keynote/}). 
}
}
\label{fig4layers} 
\end{figure}

\begin{figure}[bth]
\centering \includegraphics[scale=0.60,angle=90]{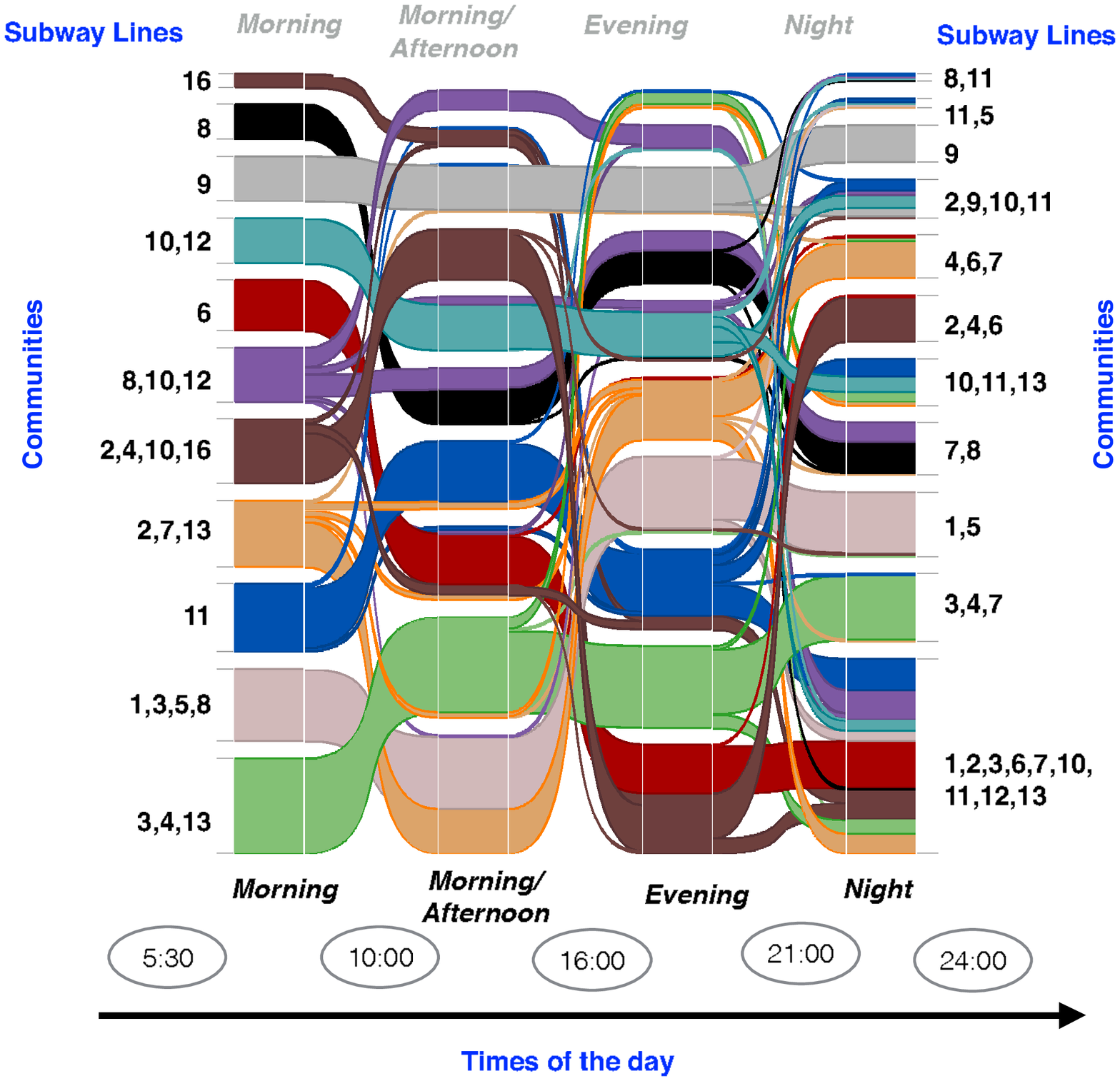} \caption{{Changes in communities over one working day. } 
\textcolor{black}{ Using the same day as in Fig. \ref{fig4layers}, this figure illustrates the changes in the communities over the day. The colours indicate the station's different community classes. The x-axis is indexed by the four time periods (morning, morning/afternoon, evening and night) in a day. The numbers on the vertical axes denote the subway lines (as described in Tab. S1) in the relevant communities. There are 11 communities in the morning and night periods, but only 10 in the other periods. Clearly, the subway lines associated with the city centre tend to construct communities with other subway lines. For example, in the morning, 5 of the 6 communities with more than 2 subway lines include subway lines 10 and 13, which are associated with the city centre.
Furthermore, the snapshots of communities in any two adjacent periods differ from each other; the differences are especially strong between the evening and night.
}
}
\label{figAllu} 
\end{figure}

\begin{figure}[th]
\centering \includegraphics[width=0.35\textwidth,angle=90]{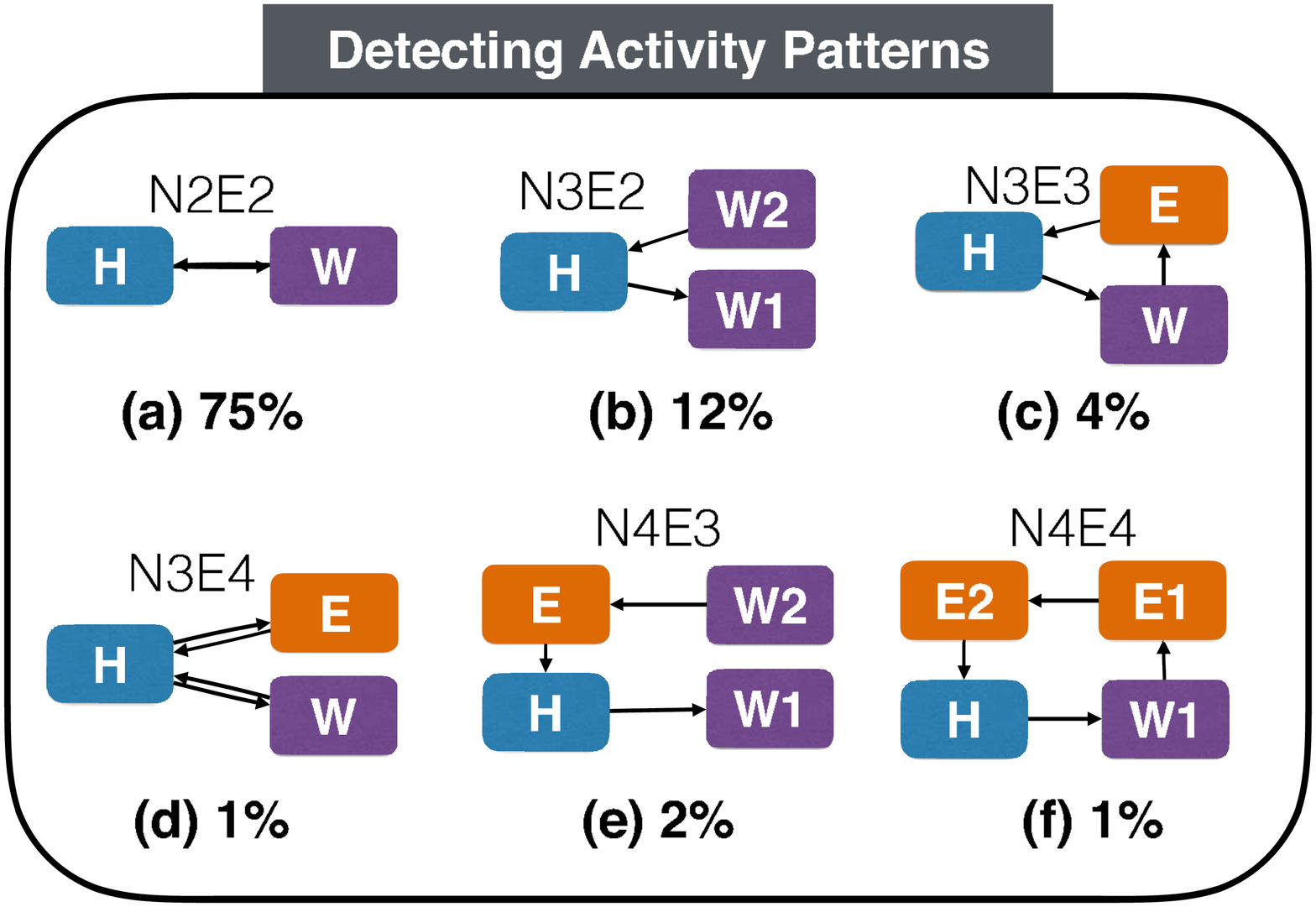} \caption{{Overview of Activity Patterns} Each activity type is marked as either H (Home), W (Workplace) or E (Entertainment). W1 and W2 denote workplace activities. Although an individual may have only one job, he/she may also go to and from work through different stations. Similarly, E1 and E2 denote entertainment activities.
\textcolor{black}{
 (a) We label each activity pattern according to the number of nodes and edges. For example, the activity pattern N2E2 denotes an activity pattern with two {N}odes and two {E}dges.
The percentages beside each pattern show the ratio of the number of individuals engaged in this kind of activity pattern to the total number of individuals.
The data on individuals' digital traces can be divided into six main activity patterns that each account for more than 1\% of the sample. }
}
\label{figActivityPatterns} 
\end{figure}

\begin{figure}[th]
\centering \includegraphics[scale=0.60,angle=90]{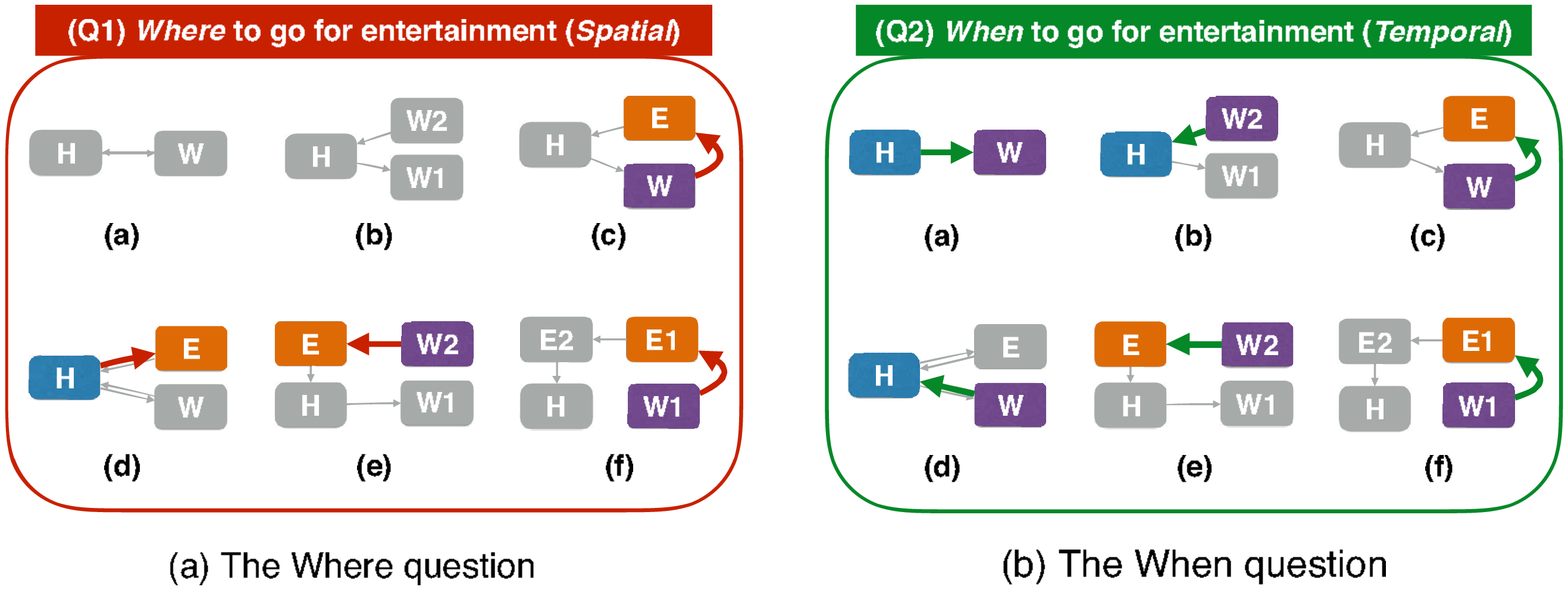}
\caption{{ Activity Patterns.} The red and blue edges denote where and when passengers go for entertainment in the evening. Note that in the calculation of the when they travel for entertainment, we include the trip from the workplace to home, due to its similarity to the travel from the workplace to an entertainment venue. Both trips are the first activities after work. Taking the time of leaving work as the beginning time, we use these trips to estimate wait times.}
\label{fig:subfig}
\end{figure}

\section*{Results}
 \subsection*{Mining community snapshots for activity patterns}
To characterise the properties of community snapshots, we start by analysing the daily community snapshots (which are the aggregate of the community snapshots of the four periods) for the studied month. Of the 18 working days, the daily community snapshots are the same for 17 days; the exception is the first working day after a holiday (Qing-ming Day).
The common daily community snapshots for each period are shown in Fig. \ref{fig4layers}. Each period is represented by a snapshot; the changing colours of the station nodes show the communities joining and splitting over the four periods.
\textcolor{black}{
To be more specific, Fig. \ref{figAllu} shows the community snapshots of the subway system that appear to represent 94\% of working days. The subway lines that make up each community are displayed. The relationships between subway lines and administrative divisions are given in Tab. S1. 
The subway lines that pass through the city centre are likely to construct communities with other subway lines that are associated with residential, tourist and business areas. For example, in the morning, five of the six communities with more than two subway lines include subway lines 10 and 13, which cross the city centre, as described in Tab. S1.
Furthermore, the number of stations in each community class is relatively stable, with one more or less between adjacent periods. However, the snapshots of the communities over all four periods (morning, morning/afternoon, evening and night) do not conform to this pattern, as stations in the last period join other communities.
The mixing of communities is particularly obvious in the evening. For example, the black community (subway line 8) in the evening combines with the purple community (subway line 8, 10 and 12).
Thus, no single community snapshot accurately shows the interactions between stations throughout a working day. 
The changes reveal passengers' various activities, and the decisions they make regarding where and when to travel. }

Thus we further study passengers' activities. 
Due to the relatively small passenger volume in the night, as shown in Fig. S2, we mainly study evening behaviour; a snapshot of evening communities is shown in Fig. \ref{fig4layers}, which also has additional information on administrative divisions.
\textcolor{black}{
Our analysis focuses on workers who start the first trip before 10:00 and make other trips after 17:00, as their activity patterns reflect regular behaviour. The high-order structure of their activity patterns is described in Fig. \ref{figActivityPatterns}. 
We label each activity pattern by their number of nodes and edges. For example, the pattern N2E2 denotes an activity pattern with two {N}odes and two {E}dges. 
We find that the activity pattern of N2E2 is the most common, followed by N3E2 (with three nodes and two edges) and others. 
Each activity pattern describes a daily activity scenario. 
For example, the pattern N3E2 describes an individual who goes from home H to workplace W1, then back home after work from place W2.
The activity patterns in the subway system show that most (85\%) individuals tend to go home after work, following pattern N2E2. Only a few people (around 15\%) go to other places after work. 
We conclude that an activity with more edges has less probability of occurring.
This is consistent with previous studies of other cities, such as Singapore \cite{jiang2015activity}, Paris \cite{schneider2013unravelling} and Chicago \cite{schneider2013unravelling}. 
}

 \subsection*{Exploring the where and when dimensions}

The community snapshot approach can improve our ability to measure the dynamics of mobility flows. However, the {spatial-temporal} models that drive the mobility dynamics are still unclear. 
\textcolor{black}{
From morning to afternoon, individuals' spatial and temporal activities are stable, as they have fixed homes and workplaces. However, in the evening, travel to entertainment places is unstable. Thus we further analyse the evening activities by exploring the where and when dimensions, as  shown by  the coloured edges shown in Fig. \ref{fig:subfig} (a) and Fig. \ref{fig:subfig} (b), respectively . Next, we describe them separately.
}

\begin{figure}[t] 
\centering \includegraphics[scale=0.60,angle=270]{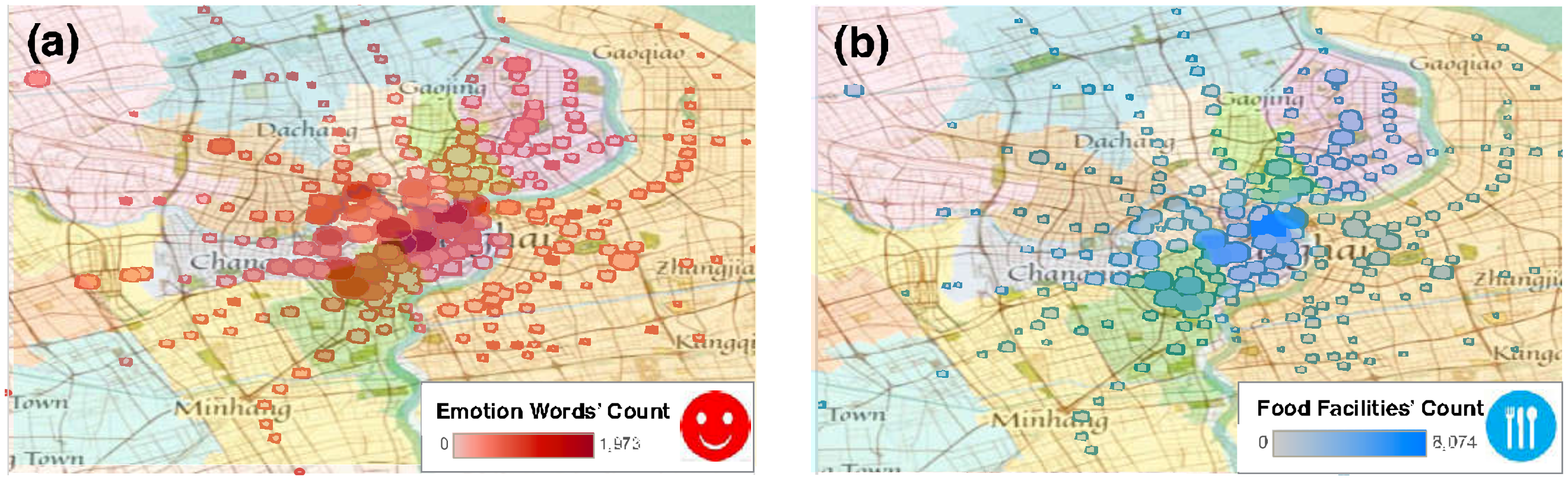} 
\caption{{Left: Map of emotions. Right: Map of food-related facilities } 
(a) This is a map of emotions. The redder the colour, the more positive the emotions in the station's environment. (b) This is a map of food facilities. The bluer the colour, the greater the density of food-related facilities.
The spatial map was created using OpenStreetMap online platform (\href{http://http://www.openstreetmap.org/}{http://http://www.openstreetmap.org/}) (© OpenStreetMap contributors) under the license of CC BY-SA (\href{http://www.openstreetmap.org/copyright}{http://www.openstreetmap.org/copyright}). More details of the licence can be found in  \href{http://creativecommons.org/licenses/by-sa/2.0/}{http://creativecommons.org/licenses/by-sa/2.0/}.
Line graphs were drawn using Tableau Software for Desktop version 9.2.15 (\href{https://www.tableau.com/zh-cn/support/releases/9.2.15}{https://www.tableau.com/zh-cn/support/releases/9.2.15}).
The layouts were modified with Keynote version 6.6.2 (\href{http://www.apple.com/keynote/}{http://www.apple.com/keynote/}). 
} 
\label{figMapEmotion}
\end{figure}

\begin{figure}[th]
\centering 
\includegraphics[scale=0.50,angle=270]{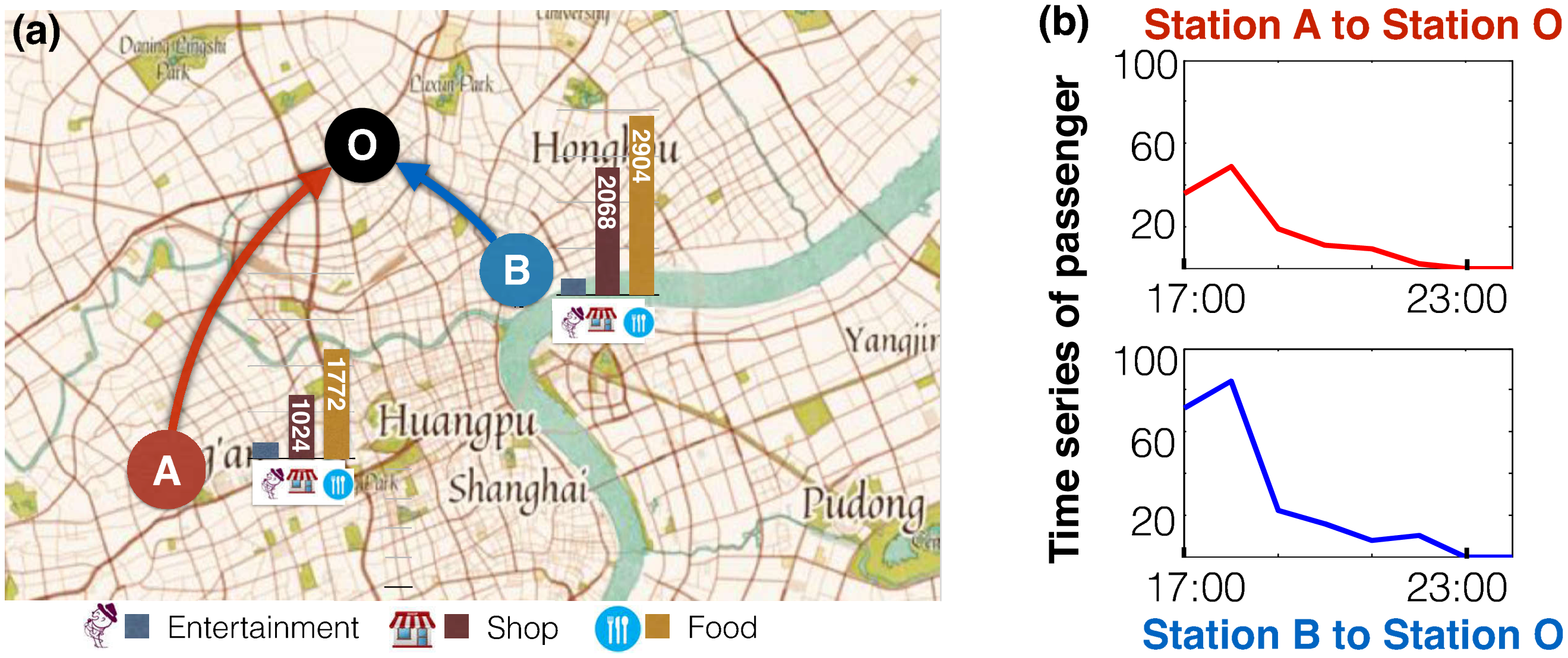} 
\caption{{Left: An example of two mobility flows. Right: Temporal series of mobility flows } (a) The two mobility flows have the same destination but two different origins with different environments. (b) The relevant temporal flows in the evening are shown with the same colours corresponding to the two samples of mobility flows.
The spatial map was created using OpenStreetMap online platform (\href{http://http://www.openstreetmap.org/}{http://http://www.openstreetmap.org/}) (© OpenStreetMap contributors) under the license of CC BY-SA (\href{http://www.openstreetmap.org/copyright}{http://www.openstreetmap.org/copyright}). More details of the licence can be found in  \href{http://creativecommons.org/licenses/by-sa/2.0/}{http://creativecommons.org/licenses/by-sa/2.0/}.
Line graphs were drawn using Tableau Software for Desktop version 9.2.15 (\href{https://www.tableau.com/zh-cn/support/releases/9.2.15}{https://www.tableau.com/zh-cn/support/releases/9.2.15}).
The waves figures in (b) are drawn by Matlab version 2016a (\href{https://www.mathworks.com/}{http://www.apple.com/keynote/}).
The layouts were modified with Keynote version 6.6.2 (\href{http://www.apple.com/keynote/}{http://www.apple.com/keynote/}). 
}
\label{fig_example2waves}
\end{figure}

\subsubsection*{Exploring the where dimension }
First, we analyse the emotional strength of each station in the eight administrative divisions of Shanghai. Specifically, 
the distribution of emotion for each station is described in Fig. \ref{figMapEmotion} (a) which shows a heterogeneous spatial distribution. 
The colours vary from blue to red. For each station, the redder the colour, the more positive the emotion.
We find that the stations in the city centre around the 'Jing'an' and 'Huangpu' divisions have higher positive emotions than the outer divisions. 

Then, we apply the microblog-based spatial model to different groups of administrative divisions, as shown in Tab. \ref{tableGuassin}. The correlation weight is a value between 0 and 1. The larger the correlation, the more accurately the model describes the reality.
We find that the divisions near the city centre have higher correlations than divisions on the edge of the city. Specifically, 
the model of the stations in the city centre has the highest correlation with reality. 
In contrast, the gravity mobility model \cite{Alonso1976} is much less accurate, with correlations around 0.1 in all three groups shown in Tab. \ref{tableGuassin}. 
\textcolor{black}{
These results suggest that considering the heterogeneous spatial popularity of stations in addition to the influence of distance can improve the description of individuals' spatial mobility flows.}

\begin{table}[!ht]
\centering \caption{{Correlations of the real mobility flows between stations with the predictions of the spatial model. } There are two groups ('Centre' and 'Outer') of administrative divisions. 'Centre' divisions are divisions 1 to 4 in Fig. \ref{fig4layers}, whereas 'Outer' divisions are 5 to 8 in Fig. \ref{fig4layers}. A third group, which combines the two groups, is labelled 'Centre\&Outer'.
}
\label{tableGuassin} %
\begin{tabular}{lll}
\toprule
 & Pearson's correlations & 95\% confidence interval of Pearson's correlations \\ \midrule
Centre\&Outer & 0.32  & 0.31 $\sim$ 0.32     \\
Centre & 0.35  & 0.34 $\sim$ 0.36     \\
Outer & 0.30  & 0.29 $\sim$ 0.31     \\ \bottomrule
\end{tabular}
\end{table}

\subsubsection*{Exploring the when dimension }
\textcolor{black}{
To investigate the temporal mobility flow, especially in the evening, we further explore the underlying temporal model by considering the influence of the environment. } 
The model examines the influence of different types of places by varying the facilities, such as the food-related facilities shown in Fig. \ref{figMapEmotion} (b). 
Fig. \ref{fig_example2waves} presents two samples of mobility flows that each have two origins and one destination. 
The two origins are in different environments with different facilities around the stations, which may play a role in individuals' temporal decision making.

\begin{figure}[h]
\centering \includegraphics[scale=0.50,angle=90]{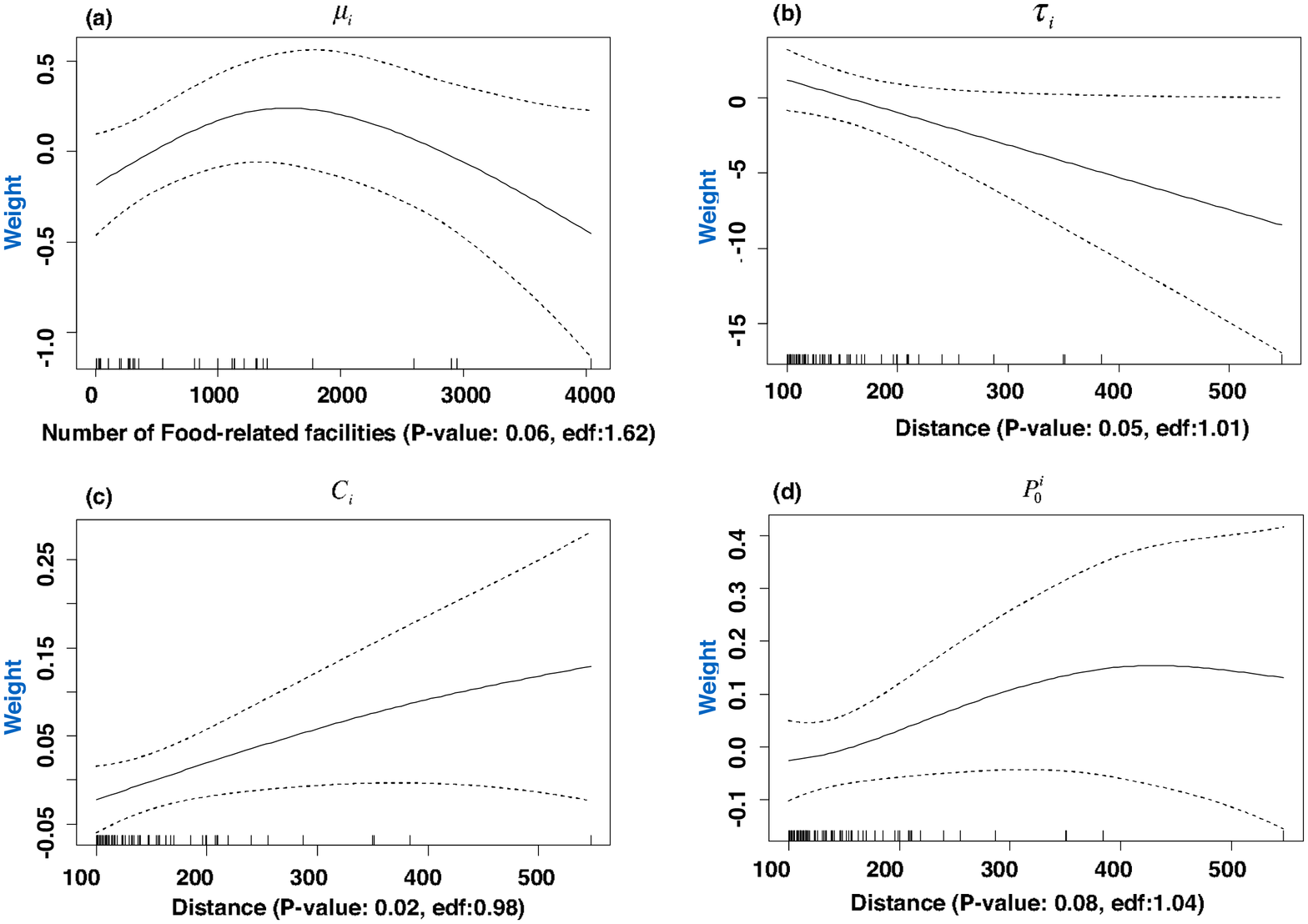} \caption{{Relationship between parameters and model smooth terms.} Four parameters are analysed here. Specifically, $\mu_i$ denotes individuals' perception of travelling discomfort. A higher $\mu_i$ indicates a higher tendency of individuals to avoid travelling discomfort.
$\tau$ represents passengers' perceptions of delay cost. A higher $\tau$ indicates a higher tendency of individuals to begin their travel as quickly as possible. 
$C_i$ is a constant, denoting the basic dynamic of individuals' temporal travelling.
$P_0^i$ refers to the initial probability of individuals' temporal travelling at the starting time. 
The mean value of the model smooth term is plotted with solid lines and the confidence intervals are plotted with dashed lines. 
In addition, on the x-axis, the $P$-value and edf of the significance of the model smooth terms are shown. 
 (a) The nonlinear relationship between $\mu_i$ and the number of food-related facilities, with the peak at (1563,0.3578). (b-d) The nearly linear relationship among parameters ($\tau_i$, $C_i$ and $P_0^i$ ) and distance. }
\label{figR_para4} 
\end{figure}

To identify the meaningful characteristics of facilities, 
we first use training samples (80\% of all of the samples)to analyse the correlations between the spatial facilities and the parameters in the temporal model. The samples are shown in the x-axis of Fig. \ref{figR_para4}. 
A $p$-value for a model's smooth term that is less than or equal to 5\% or 1\% indicates that the chosen smooth term is significant. 
\textcolor{black}{ Moreover, the {e}stimated {d}egree of {f}reedom (edf) for the smooth terms' significance is estimated. } 
When the value of the edf is far from 1, the smooth item tends to reflect a nonlinear relationship. 
In this study, keeping only the significant features ($p$-value $<$ 0.1), we find that only the food-related feature, with $P$-value 0.06, has an obvious correlation with $\mu$. The edf is far from 1, indicating a nonlinear relationship between individuals' perceptions of travelling discomfort and the presence of food-related facilities, as shown in Fig. \ref{figR_para4}. 
Individuals near stations with a higher $\mu$ tend to have lower travelling discomfort when the number of food-related facilities is near 1563.
As the number of food-related facilities increases, $\mu$ first increases, then decreases, indicating that a moderate number of food-related facilities has the greatest influence.
As for the perception ($\tau$) of delay cost, distance plays an obvious role, with a $p$-value of $<$ 0.05. $\tau$'s edf of the smooth term on distance is near 1, indicating a linear relationship between perception of delay cost and distance. Specifically, as shown in Fig. \ref{figR_para4}, individuals near stations with higher $\tau$ values tend to start their travel earlier. 
The other two parameters ($C$ and $P_0$) are only related to distance, with which they have a positive linear relationship. 

In addition, applying the correlations learned by the training samples, we use the testing samples (20\% of all samples) to examine whether it is possible to simulate the temporal flows with the inferred parameters. 
There are 17 pairs of stations in the testing sample. We mainly study their temporal flows in the evening after work (from 17:00 to 24:00). The simulation result is shown in Fig. S1. 
 As we see, the simulation based on the temporal model driven by the desire to balance travelling comfort and delay cost is much closer to the real temporal flows. 
 \textcolor{black}{
 This result shows that our temporal model describes to some extent the real temporal mobility flows, and can take into account the influences of the environment on individuals' perceptions.}

\section*{Discussion}
To study the dynamics of mobility flows in urban transport it is important to estimate the spatial and temporal interactions of human mobility flows between stations. Understanding these flows helps planners to evaluate traffic congestion \cite{ceapa2012avoiding}, and avoid the dangers of overcrowding \cite{pearl2015crowd}). 
Taking the subway system of Shanghai as a case study, we used community snapshots to investigate the {variability of spatial-temporal} mobility flows on working days. The results show that the flows change at different times of day,  but the patterns are similar on each working day. 
Recognising these dynamics will help to predict human movements between stations on spatial-temporal scales, which may help planners to  efficiently schedule subway cars.

To further understand the models, we investigate the {high-order structure of activity patterns} in both the temporal and spatial dimensions. Specifically, 
we use the patterns in passengers' travelling activities to determine passengers' lifestyle needs and then investigate (1) the {where (spatial)} dimension of mobility flows between subway stations by correlating microblog topics with spatial popularity and 
 (2) the {when (temporal)} dimension of individual schedules of activities by correlating spatial facilities and travelling distance with the perceptions of travelling discomfort and delay cost.
We argue that correlations between the stations and their environments may to some extent be explained by urban catalyst theory (first proposed in 1960 \cite{jacobs1961death} and fully developed in 2000 \cite{davis2009urban}).
According to this theory, railway stations can act as urban catalysts and have positive effects on their surroundings, but are also influenced by their surroundings \cite{geng2009effect,jiwei2006railway,papa2008rail}. Therefore, a boom in living/business buildings or popular locations around a subway station can increase its traffic flows, as appears to be the case in Shanghai\cite{jiwei2006railway}, Seoul\cite{bae2003impact}, Kun Dae Yeok\cite{lee1997accessibility} and Toronto \cite{dewees1976effect}.
Note that the nonlinear relationship between $\mu_i$ and food-related facilities, shown in Fig. \ref{figR_para4}, indicate that although the overall influence of food-related facilities is related to the quantity of facilities, more is not necessarily better. Once the number of food-related facilities in a specified spatial area exceeds a threshold, which may vary with the size of the area, many small but low-level food-related facilities may appear, thus decreasing the total influence of food-related facilities.
This study has some limitations. 
Community snapshots of the Shanghai subway system are used as a case study, and the dataset contains information for only one month. Other patterns may be found with data covering a longer time period. However, the same methods analysis of could be used. In this study, we focus on community snapshots of the subway system; future studies could consider the similarity and differences of the patterns in other public transit systems.
 \textcolor{black}{
{Furthermore}, in this study we only analyse the correlations of individual spatial-temporal perceptions with the environment. The causal relationships should be examined in future studies. In this study we characterise the environment by counting the number of various kinds of facilities, but do not consider their scales. Thus we do not sufficiently capture the correlation with entertainment and shopping facilities, as their effect on individuals' perceptions cannot be measured by simply counting the facilities. Better measures are needed to further analyse their effect.
}

\section*{Methods}
 \subsection*{Data}
To study our tasks, we use the travel smart card dataset from Shanghai, China for the month of April 2015. Items (such as card number, transaction date, transaction time, station name and transaction amount) are recorded by the subway system. The dataset contains information about 313 subway stations, 11 million individuals and around 120 million trips. We define the morning period as from 5:30 to 09:59, and the evening period from 16:00 to 20:59. Both periods are approximately 5 hours. The two time periods between these periods are labelled morning/afternoon and night. A trip is only counted as belonging to a given period if both the starting time and end time occur within the same period. 
More details can be found in the Supplement $\S 2.1$.

Baidu APIs \cite{baidu_Shanghai} is used to identify the 2 million spatial facilities within 1 km of a subway station. The spatial facilities are assigned to one of three categories (entertainment, shopping and food) according to their business type. We also use the Weibo APIs\cite{Weibo_Shanghai} to identify 1 million microblogs generated near the stations.
To identify the topics of these microblogs, we use the Topic Expertise Model, an improved Latent Dirichlet allocation (LDA) topic model for the detection of keywords \cite{yang2013cqarank}; we use the groups of microblogs generated near each station as files.
Finally, we determine the distribution of topics across the subway system using the word distribution in each topic. See Supplement $\S 2.1$ for details.

 \subsection*{Mining community snapshots }
To address the technical challenges of the community snapshot analysis for each period, we use community detection methods. The community snapshot analysis constitutes four steps. For each day, there are different temporal scales of flow matrices, representing the directed mobility flows between any two stations at a certain temporal scale. In this study, we focus on the temporal scale of four periods in a day. Through community detection and consensus analysis of each period, we can build a matrix of community identification for each station, producing a community matrix for each period. With the help of clustering methods, we can identify the categories of these community snapshots. See Supplement $\S 2.2$ for details.

 \subsection*{Spatial and temporal models for the where and when dimensions}
\textcolor{black}{We adopt a dual-perspective on passengers' activity patterns to investigate the spatial and temporal dimensions of individuals travelling decision making. Specifically, we examine where and when individuals go for entertainment in the evening, by introducing an individual-based model.}
Specifically, we use microblogs, which reflect individuals' relevant perceptions of popularity and distance, to determine the {heterogeneous spatial popularity} $X_j^i$ of each topic at every station. This results in a {spatial} model based on the location-based mobility model. 

To study the {temporal} model of individual schedules of activities, we use the difference equations of temporal flows to model the dynamic evolution of balancing travelling comfort and delay cost, which are correlated with environment and distance. 
Let $Y_{t}^{i}$ denote the travelling volume from the $i$-th station at time $t$. 
Thus $Y_{t}^{i}$ is modelled as  $
Y_{t}^{i}=N_{i}(\tau_i \Delta t - \mu_i Y_{t-1}^i + C_i)$. 
Here  let $\mu_{i}$ refers to individuals' perceptions of the travelling discomfort in the $i$-th station, correlated with the nearby economic/social buildings.  $\tau_i$ refer to individuals' perceptions of the delay cost, correlated with the distances between stations. $N_i$ denotes the total volume of passengers in station $i$ in the evening. Thus individual's temporal choice at time $t$ has a probability of $p_t^i=Y_{t}^{i}/N_{i}$.
In addition, these parameters (such as $\mu_i$ and $\tau_i$ ) are further correlated with features (such as economic/social buildings and distances) by the generalised additive model. 
See Supplement $\S 2.3$ for more details.


\section*{Acknowledgements (not compulsory)}

The authors are grateful to Dr Yongjian Yang  for his valuable support of high performance computing cluster.

\section*{Author contributions statement}
J.L. , B.Y. and Z.D. defined the problem. Z.D. J.L. and Y.B. discussed the method. Z.D. implemented and refined the method and conducted the experiments.All authors reviewed the manuscript.

\section*{Additional information}

Competing financial interests: The authors declare no competing financial interests.
Supplement can be found in URL: https://www.overleaf.com/read/qbdzxhsgfmbm

\end{document}


\flushbottom
\maketitle
\thispagestyle{empty}

\section*{

\section*{Methods}
 \subsection*{Data}
\subsubsection*{Traffic data}
To study our tasks, we employ the travel smart card dataset (collected by Shanghai Public Transportation Card Co.Ltd) in Shanghai, China during 30 days in April 2015.  Items (such as card number, transaction date, transaction time, station name, and transaction amount) are recorded by the subway system. The dataset contains information of 313 subway stations, 11 million individuals and 123 million events of trips. If a person checks in or out the subway stations with smart card, when and where this event will be recorded automatically by the subway system. A case study of this dataset can be found in the dataset website \cite{coltd2015}.
Previous studies \cite{long2015finding,chatterjee2015studies} focus on the existence of different community snapshots for the weekdays and weekends. However, in reality, there mye be still holidays in a month, which may result in different human patterns triggered by abnormal passengers' mobility behaviors. In our study, there are two holidays needed to be considered. One is Qing-ming Day which lasts from April 4 to April 6. Another is Labor Day for three days off from May 1. 
We consider the contact matrix to reveal the community snapshots in working days without holidays. Specifically, the \textbf{flow matrix} of stations is collected for each hour during a day by identifying passengers' starting and ending subway stations for every trip within the time range of Hour, organized by the period contact matrices of stations.  There are mainly two periods of peak hour (e.g., morning peak hour and evening peak hour) in the subway system of Shanghai \cite{sun2015research}. Thus we mainly consider the period community snapshots of four  periods in a day, shown in Tab.\ref{table2}. 
The total passenger volume for each hour in a working day  is shown  in Fig. \ref{figCurvePassenger}. There are two traffic peaks (morning and evening). To cover the morning and evening peaks, the period of morning is set from 5:30 to 09:59, and the evening from 16:00 to 20:59. Both are for around  5 hours. As for the left two time gaps, they are assigned to the periods of morning/afternoon and night.
For each period, only when the starting time and end time of a trip all belong to the time range of a period, the trip is used for this period.

\subsubsection*{Spatial facilities' data}
We use the Baidu APIs \cite{baidu_Shanghai} to get the spatial facilities around each station within 1km. There are more than 2 million spatial facilities collected with detail business information (such as name, business type, address and so on). 
To reflect the living comfort around each station \cite{liu2014residential}, different kinds of spatial facilities are categoried into three types (entertainment,  shopping, and food)  according to their business type and then taken as features to quantify individuals' perception of traveling comfort. 
Specifically, the type of entertainment denotes facilities' business type of entertainment.
The type of shopping represents facilities' business type of shopping.
Moreover, the type of food represents facilities' business type of food, but not including the related small business (such as the cake shop, teahouse, and bar).

\subsubsection*{Microblog data}
Corresponding to the studied period of traffic data, we use the Weibo APIs\cite{Weibo_Shanghai} to get the nearby microblogs around each station.  There are around 1 million microblogs collected, containing around 15 thousand nominal keywords filtered by the Chinese word segmentation methods \cite{nlpcn}.  
As for the emotion strength of each keyword,  we map them into the emotion dictionary of Chinese words \cite{yu2008constructing}.
Furthermore,  to find the topics behind these microblogs, the Topic Expertise Model, an improved Latent Dirichlet allocation (LDA) topic model for the detection of keywords, is used here \cite{yang2013cqarank}, taking the groups of microblogs around stations as files.
Finally, we can get the topic distribution in each station and the word distribution in each topic.

\begin{figure}[h]
\centering \includegraphics[scale=0.3]{CurvePassenger} \caption{\textbf{The temporal flows in working days.} The total passenger volume of all working days  is shown  for each hour. There are two traffic peaks (morning and evening). To cover the morning and evening peaks, the period of morning is chosen from 5:30 to 09:59, and the evening from 16:00 to 20:59. Both are for around  5 hours. As for the left two time gaps, they are assigned to the periods of morning/ and night.
}
\label{figCurvePassenger} 
\end{figure}

\begin{table}[!ht]
\centering \caption{Periods involved in a working day}
\label{table2} %
\begin{tabular}{ll}
\hline 
\textbf{Period Name}  & \textbf{Period Section}\tabularnewline
\hline 
Morning & 5:30-09:59 a.m.\tabularnewline
Morning/Afternoon  & 10:00-15:59\tabularnewline
Evening  & 16:00-20:59\tabularnewline
Night & 21:00-23:59\tabularnewline
\hline 
\end{tabular}
\end{table}

\begin{algorithm}
\caption{Key steps in the analysis of community snapshots } \label{alg1}
\textbf{Input}: Flow matrices of day and periods for each day

\textbf{Output}: Categories of daily community snapshots
\begin{itemize}
\item \textbf{Step} \textbf{1: Detecting the community snapshots for each
period in a day}
\begin{itemize}
\item \textbf{For} each working day $d_{i}$ \textbf{do}
\begin{enumerate}
\item \textbf{For} each period $pd_{i}$ \textbf{do}
\begin{enumerate}
\item Detect communities based on period flow matrix
\item Take the snapshot of communities as period community snapshot
\end{enumerate}
\item Take the combination of all period community snapshots as periods-day
community snapshot
\end{enumerate}
\end{itemize}

\item \textbf{{Step} {2: Making consensus clustering for the community snapshots of all periods}}
\begin{itemize}
\item Update each stations' periods-day community identification in periods-day community
patterns 
\end{itemize}
\item \textbf{Step} \textbf{3: Measuring variability of community snapshots}
\begin{itemize}
\item Compute the correlation matrix for periods-day community snapshots
\end{itemize}
\item \textbf{Step 4: Clustering the community snapshots based on the correlation
matrices}
\begin{itemize}
\item Cluster the periods-day community snapshots into different categories
\end{itemize}
\end{itemize}
\end{algorithm}

 \subsection*{Community snapshots mining}
To perform the community snapshot analysis for each periods in a day, we address the technical challenges through community detection methods. As depicted in Alg. \ref{alg1}, the community snapshot analysis constitutes 4 steps. For each day, there are different temporal scales of flow matrices, representing the directed mobility flows between any two stations in certain temporal scale. Here, we mainly consider the temporal scale of periods in a day. Thus a day has one form of expressions, with different periods of flow matrices. Through community detection and consensus analysis on them, we can get the matrix of community identification for each station, as community matrix for periods-day community snapshot . With the help of clustering methods, we can get the categories of periods-day community snapshots.  Next, we describe these steps in more details.

\subsubsection*{Community snapshots detecting and consensus clustering} 
In this study, we use community detection method (CDM) to perform the analysis of community snapshots in UTN with the period contact matrix of stations as input. CDM offers the capability to divide the network into groups with dense inter-connections and sparser intra-connections, to probe the underlying structure of complex networks and find useful information from them \cite{sobolevsky2014general}. CDM is well suited for our study to examine the human mobility patterns in UTN by observing the forming and changing process of major communities.

There are many types of CDM techniques, such as greedy agglomerative optimization by Newman \cite{PhysRevE.69.066133} and faster Clauset-Newman\_Moore heuristic \cite{PhysRevE.70.066111} and so on. The major distinguished difference between them lies in the objective function, which is used for partitioning.  Most CDMs are based on modularity as objective function\cite{sobolevsky2014general}. There are also some other objective functions used to motivate the CDMs, such as description code length, block model likelihood measure, and surprise. Our study aims to find the community snapshot in UTN in general, which is more suitable to the widely used objective function. Therefore, we employ a high-quality modularity based CDM (called Combo)\cite{sobolevsky2014general} as the adopted analysis technique for day community snapshots. As for the analysis of  community snapshots, the regularity of mobility patterns in the subway system varies according to different temporal scales in a day \cite{zhong2016variability}. Thus, the kind of hierarchical CDMs with modularity optimization is suitable to detect the communities, with supplying different angularities of community identification. 
Our study aims to find suitable community identifications for each stations in every period. Therefore, a high-qualify Hierarchical CDM (Louvain method \cite{rosvall2008maps}) is employed for the period community analysis. 

As for the consensus clustering, the consensus of temporal networks has been well studied to track the evolution of node in the dynamic communities\cite{lancichinetti2012consensus}. Many community detection algorithms (e.g., Louvain method ) have been studied combined with consensus clustering. Thus we choose the consensus clustering method \cite{lancichinetti2012consensus}, which have been studied combined with many algorithms (e.g., Louvain method) for the consensus clustering for the community snapshots of all days. 

\subsubsection*{Variability measure of daily community snapshots.}
The dynamics of daily community snapshots during the working days (without holidays) in  the studied month is represented as a succession of snapshots of periods-day community snapshots by the above community detection methods. For each form of community snapshots, to identify the regularity and find whether two days have the same community snapshot, we need to measure the variability between them. Regarding human mobility patterns in the subway system, the degree of regularity between any two days is measured through the correlation of their temporal vectors of features from the normalized covariance matrix \cite{zhong2016variability}. Following this way, we use the nominal correlation to measure variability. There are many types of nominal correlation, such as Pearson contingency coefficient, Spearman Rho, and Cohen\textquoteright s Kappa. The daily community snapshots show variability, which is more suitable to use the nominal correlation based on information theory concepts. Thus we take the probability-distribution based Pearson contingency coefficient as the nominal correlation in this work.

\subsubsection*{Clustering the daily community snapshots}
With the above variability matrix of daily community snapshots, the next step is to cluster the patterns, to identify the different categories of daily community snapshots. One disadvantage of the existing popular clustering algorithms (e.g., hierarchical clustering algorithms and k-means algorithms) is that they are largely heuristic and not based on formal models. While model-based clustering is an alternative \cite{banfield1993model,melnykov2010finite}, initialized by hierarchical clustering for parameterized Gaussian mixture models. Therefore, we employ the model-based clustering method to cluster the daily patterns. 

 \subsection*{Spatial and temporal models analysis of the where and when questions}
\todo{ the algorithm frame are used to introduce the steps of methods vivid.  }
\textcolor{black}{We further take a dual-perspective view on passengers' activity patterns to investigate the models of individuals traveling decision making from both spatial and temporal scales by answering where and when individuals go for entertainment in the evening.}
In this respect, we introduce an individual-based model, as described in Fig. \ref{Alg_framework}. 
Specifically,  by using the microblogs to reflect the \textbf{heterogeneous spatial popularity} $X_j^i$  (which attracts individuals in station $i$  by the $j$-th topic to spend their vacation ) for each topic in every station, the \textbf{spatial} model is studied through a location-based mobility model with considering individuals' relevant perceptions of popularity and distance .  
Moreover, the \textbf{temporal} model of individual time schedules of activities is also studied by difference equations of temporal flows with modeling the dynamic evolving process of balancing traveling comfort and delay cost (which are correlated with environment and distance). Next, we describe them in more details. Next, we describe them in more details.
\begin{figure}[h]
\centering \includegraphics[scale=0.35]{Alg_framework} \caption{\textbf{The proposed model of individuals traveling decision making.} We consider an individuals traveling decision making in dual-perspective view  of spatial and temporal models. 
In doing so, two models are constructed to describe where and when an individual travel under various social influence with considering individuals' perceptions. 
\textcolor{black}{
In the \textbf{spatial} model of the \textbf{where} question, we apply the microblog topics to characterize the spatial popularity (denoted by $X_j^i$). Then to get the spatial attraction, the model of spatial decision making is constructed by combining individuals' popularity perception  (denoted by $\theta_j$) with  spatial popularity.
Here, $p^{s\rightarrow i}$ denotes to the spatial probability, which is the probability for an individual choosing station $i$ as destination.
And we also integrate the \textbf{temporal} model of  the  \textbf{when} question with the influence of spatial facilities by characterizing  people's perceptions of traveling discomfort   (denoted by $\tau_i$)  and delay cost   (denoted by $\mu_i$). Here $p_t^i$  refers to the temporal probability, which is the probability for an individual choosing time $t$ for leaving. }
}
\label{Alg_framework} 
\end{figure}

\subsubsection*{Spatial model analysis  to explore the where question}
Let  $X_{j}^{i}$ denote the popularity of  the $j$-th topic in the $i$-th station. To be simple, we take the appearing probabilities of words (which are nearby the $i$-th station) in the $j$-th topic as $X_{j}^{i}$. Mathematically,  $X_{j}^{i}$ can be written as the following form:
\begin{equation}
X_{j}^{i}=\sum_{\omega=1}^{N_{i}} p_{j}^{\omega}
\end{equation}
Let $p_{j}^{w}$ be the probability of the $w$-th word in the $j$-th topic. $N_{i}$ refers to the size of words nearby the $i$-th station.  

Then we simply assume that the attraction of a destination station is inversely proportional to  the spatial popularity  of  topics. Specifically, the relative attraction $A^{s \rightarrow i}$ of the destination station $i$ to passengers at the origin station $s$  is described as :
\begin{equation}
A^{s \rightarrow i}=o_{i}\frac{1}{1+e^{-(\sum _{j=1}^{M} \Theta _{j} X_{j}^{i}  + \Theta _{d} d_{sj}+ \epsilon ) }}
\end{equation}
Let $\epsilon$ and $o_{i}$  denote the relative residual error and the total opportunities of destination $i$  respectively. 
  $\Theta_{d}$ refers the normalized impact of distance $d_{sj}$ between station $s$ and $j$. 
 Suppose there are $M$ categories of topics,   and   let   $\Theta_{j}$ describe the impact of $X_{j}^{i}$ as the normalized emotion strength $E_{j}$ of the $j$-th topic. 
  Both   $\Theta_{d}$  and   $\Theta_{j}$  are normalized with the upper bound $ub$ and the lower bound $lb$.  For instance, mathematically,  $\Theta_{j}$  can be written as the following form:
\begin{equation}
\Theta _{j} = \frac{ E_{j}-lb  }{ub-lb}
\end{equation}
Let  $e_{k}$ be the $k$-th word's emotion strength. And  $E_{j}$  can be taken as the sum of words' emotion strength in the $j$-th topic ( which contains $D_{j}$  words):
\begin{equation}
E_{i} = \sum_{k=1}^{D_{j}} (e_{k})
\end{equation}
Further, with assuming that the traveling probability $p^{s \rightarrow i}$ from origin $s$ to destination $i$ is proportional to the attraction of $i$, $p^{s \rightarrow i}$ can be described as:
\begin{equation}
p^{s \rightarrow i}=\frac{A^{s \rightarrow i}}{\sum_{k=1}^{N_s}A^{s \rightarrow {k}}}
\end{equation}
where $N_s$ is the set of all stations, expect the $i$-th station.
If a individual in station $s$, the probability of he/she chooses the station $i$  as destination follows the  the traveling probability $p^{s \rightarrow i}$ .

\subsubsection*{ Temporal model analysis to explore the when question}
\textcolor{black}{When an individual begins his/her trip, he/she will be not only influenced by traveling comfort  and the already waiting time, but also the trip's distance and the food facilities nearby the starting station.}
Let $Y_{t}^{i}$ denote the traveling volume from the $i$-th station at time $t$. 
We assume  $Y_{t}^{i}$  is effected by individuals' perceptions of traveling discomfort and delay cost in the temporal model. Specifically, 
(1) The traveling discomfort measures the passengers' feeling of comfort and is correlated with the traffic volume \cite{de2015discomfort}. To be simple, we use $Y_{t-1}^{i}$ to represent the traveling discomfort. 
 (2)The delay cost is related the waiting time $ \Delta t $ , which is the time gap between the beginning time and the current time.

 Thus $Y_{t}^{i}$ is modeled as:
 \begin{equation}
Y_{t}^{i}=N_{i}(\tau_i \Delta t - \mu_i Y_{t-1}^i + C_i)
\end{equation}
 $\mu_{i}$  refers to individuals' perception of the traveling discomfort in the $i$-th station, correlated with the nearby economic/social buildings. And let $\tau_i$  refer  individuals' perception of the delay cost, correlated with the distances between stations.  $N_i$ denotes the total volume of passengers in station $i$ in the evening. Thus individual's temporal choice at time $t$ is with probability $p_t^i=Y_{t}^{i}/N_{i}$.
 
 Here, we use the generalized additive model to describe the correlation among  parameters (such as $\mu_i$   and  $\tau_i$ ) and features (such as economic/social buildings and distances). 
Specifically,   $E(\mu_{i})$, as the expectation of  $\mu_{i}$, is related with the features of economic/social buildings with a link function $g$  (the log functions) via the following structure:
 \begin{equation}
g(E(\mu_i))= \sum_{k=1}^{K} f_k(x_k^i)
 \end{equation}
where $x_k^i$ relates the $k$-th feature of  economic/social buildings around station $i$. And the functions $f_k$ are smooth functions. The similar is $\tau_i$ with the predictor variable of distance. 

Besides, let $P_0$  denote the initial value of  $(\tau_i \Delta t - \mu_i Y_{t-1}^i + C_i)$  at the start time.  
To be simple, the constant vales of $C_i$ and $P_0$  are assumed to correlate  with all features (such as economic/social buildings and distances) by the generalized additive model with the identity functions as the link function $g$.

\bibliography{sample}